\documentclass[journal=nanolett,manuscript=article,superscriptaddress,amsmath,amssymb,textcomp]{achemso}

\usepackage{graphicx}
\usepackage{dcolumn}
\usepackage{bm}
\usepackage{color}
\usepackage{soul}
\usepackage{amsmath}
\usepackage{changes}

\author{Freddie Hendriks}
\affiliation{Zernike Institute for Advanced Materials, University of Groningen, The Netherlands}

\author{Alexander N. Rudenko}
\affiliation{Institute for Molecules and Materials, Radboud University, Nijmegen, The Netherlands}

\author{Malte R\"osner}
\affiliation{Institute for Molecules and Materials, Radboud University, Nijmegen, The Netherlands}

\author{Marcos H. D. Guimar\~{a}es}
\affiliation{Zernike Institute for Advanced Materials, University of Groningen, The Netherlands}
\email{m.h.guimaraes@rug.nl}

\title{Electrostatic Control of Magneto-Optic Excitonic Resonances in the van der Waals Ferromagnetic Semiconductor Cr$_2$Ge$_2$Te$_6$}


\begin{document}

\date{\today}
\begin{abstract}
Two-dimensional magnetic materials exhibit strong magneto-optic effects and high tunability by electrostatic gating, making them very attractive for new magneto-photonic devices.
Here, we characterize the magneto-optic Kerr effect (MOKE) spectrum of thin Cr$_2$Ge$_2$Te$_6$ from 1.13 to 2.67 eV, and demonstrate electrostatic control over of its magnetic and magneto-optic properties.
The MOKE spectrum exhibits a strong feature around 1.43 eV which we attribute to a magnetic exchange-split excitonic state in Cr$_2$Ge$_2$Te$_6$, in agreement with \textit{ab-initio} calculations.
The gate dependence of the MOKE signals shows that the magneto-optical efficiency - rather than the saturation magnetization - is affected by electrostatic gating.
We demonstrate a modulation of the magneto-optical strength by over 1 mdeg, with some wavelengths showing a modulation of 65\% of the total magneto-optical signals, opening the door for efficient electrical control over light polarization through two-dimensional magnets.
Our findings bring forward the fundamental understanding of magneto-optic processes in two-dimensional magnets and are highly relevant for the engineering of devices which exploit excitonic resonances for electrically-tunable magneto-photonic devices.
\end{abstract}

\maketitle

\section{Introduction}
Two-dimensional (2D) magnetic materials provide a unique platform for the study and applications of magnetism in low dimensions.
Due to their low dimensionality, 2D magnets are highly susceptible to external stimuli, such as light and electric fields.
This has been used to obtain electric control over their magnetic properties, such as the Curie temperature, the magnetic anisotropy, and the magnetic exchange interaction, in both metallic \cite{Deng2018, Zheng2020, Chen2021} and semiconducting materials \cite{Wang2018a, Huang2018, Jiang2018, Verzhbitskiy2020, Zhuo2021, Zhang2020a}.
Additionally, 2D magnetic materials have demonstrated very strong magneto-optical responses\cite{Fang2018, Wu2019, KumarGudelli2019, Molina-Sanchez2020, Yang2021}, with magnetically-induced changes in light polarization (rotation or ellipticity) of several milliradians even for films of only a few atomic layers thin.
This has shown to be an ideal avenue to explore the small magnetic volumes of 2D magnets, usually unaccessible by conventional magnetometry.
Magneto-optic microscopy techniques, such as the magneto-optic Kerr effect (MOKE), are now widely used to study the magnetization and its dynamics in 2D magnets.

The strong magneto-optic response in a few 2D magnetic semiconductors have been shown to arise, in a large part, from excitonic transitions \cite{Wu2019}.
This provides a powerful pathway to combine optics and photonics with magnetism.
However, the electric control over magneto-optic parameters using excitonic resonances in 2D magnets is still largely unexplored.

Here we report the MOKE spectrum of Cr$_2$Ge$_2$Te$_6$ (CGT) in the photon energy range of 1.13 - 2.67 eV (1100 - 465 nm), showing strong fingerprints of excitonic resonances.
We show that the MOKE signals can be effectively controlled by electrostatic voltages through changes in the charge carrier density ($\Delta n$) and out-of-plane electric field ($\Delta D/\varepsilon_0$).
We find, however, that the change in saturation level of the magnetic hysteresis curves obtained by magneto-optics is dominated by a change in the strength of the magneto-optic efficiencies, rather than by a change in the saturation magnetization as often attributed in the literature.
On the other hand, the values for the coercive field and the nucleation field obtained from the magnetization curves do not depend on the strength of the magneto-optic Kerr effect, which means that MOKE can still be used to determine their gate dependence.
We observe a doubling of the coercive field from 0.3 to 0.6 mT and a $\sim $ 30\% reduction of the nucleation field from 4.0 to 2.7 mT by changing the gate-induced charge carrier density between -1.5 and 1.0 $\times 10^{13}$ cm$^{-2}$.

\section{Device Structure and Measurement Techniques}
\begin{figure} [htbp]
	\centering
	\includegraphics[width=\columnwidth]{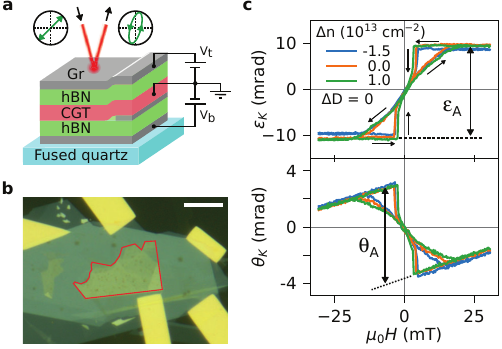}
	\caption{\textbf{MOKE in a CGT based heterostructure.}
    \textbf{a}, Illustration of the magneto-optic Kerr effect, combined with a schematic of the sample indicating the different layers and the electrical connections for gating the CGT. 
    \textbf{b}, Optical image of the sample. The CGT flake is outlined in red and the scale bar is 10 $\mu$m.
    \textbf{c}, Typical behaviour of the Kerr rotation and ellipticity during a magnetic field sweep. Measured for different values of $\Delta n$ and with $\Delta D$ = 0 at a photon energy of 1.348 eV.
    }
	\label{fig:1}
\end{figure}

Our device consists of a 10 nm thick CGT flake, encapsulated in hexagonal boron nitride (hBN), with thin graphite flakes that function as top gate, back gate, and contact electrodes for the CGT layer (see Fig. \ref{fig:1}a and b).
The whole van der Waals stack is placed on top of a transparent fused quartz substrate.
The combination of top and bottom gates allows us to independently control $\Delta n$ and $\Delta D$.
The MOKE measurements are performed in the polar geometry at 10 K.
The Kerr rotation and ellipticity are measured simultaneously by using a combination of a photo-elastic modulator (PEM) and amplified photodetector with a polarization filter \cite{Schellekens2014}.
More details about sample fabrication and the measurement setup are given in the Methods.

The Kerr rotation and ellipticity signals as a function of magnetic field show a typical magnetic hysteresis behavior, Fig. \ref{fig:1}c.
These measurements were done using a photon energy of 1.35 eV, for various values of $\Delta n$ with $\Delta D = 0$.
The shape of the hysteresis curve is typical for CGT of this thickness \cite{Wang2018a, Lohmann2019, Ostwal2020, Zhuo2021}.
It exhibits a sharp drop in magnetization before the magnetic field crosses zero, followed by a gradual change towards the new saturation level.
The external magnetic field value at which the sharp drop occurs is called the nucleation field ($H_n$).
At this field, multiple magnetic domains nucleate suddenly in the previously single domain state, which is accompanied by a reduction in the magnetization.
As the magnetic field is changed further, the multi domain state gradually evolves towards a single domain state again \cite{Lohmann2019}.
The difference in saturation level for positive and negative fields, indicated by $\varepsilon_A$ and $\theta_A$, is proportional to both the saturation magnetization and the strength of the magneto-optic Kerr effect, i.e. the magneto-optic efficiencies.
A close look to the $\Delta n$ dependence of $\varepsilon_A$ and $\theta_A$ reveals that when $\Delta n$ becomes more negative, $\Delta \varepsilon_A$ decreases while $\Delta \theta_A$ increases.
The MOKE efficiency dependences with $\Delta n$ and $\Delta D$ are investigated in more detail below.
First, we focus on the MOKE spectrum.

\section{MOKE spectrum}
\begin{figure} [htbp]
	\centering
	\includegraphics[width=0.5\columnwidth]{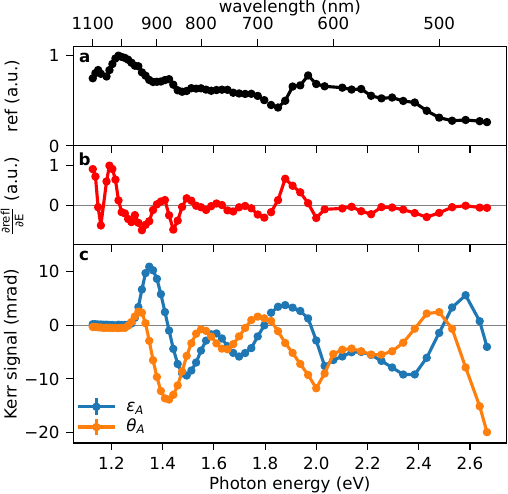}
	\caption{\textbf{MOKE spectrum of CGT.}
    reflectivity spectrum (\textbf{a}), and its derivative with respect to the photon energy (\textbf{b}), of our CGT device at $\Delta n$ = $\Delta D$ = 0.
    \textbf{c}, MOKE spectrum showing the dependence of the Kerr ellipticity ($\varepsilon_A$ - blue) and rotation ($\theta_A$ - yellow) with the incident photon energy.
    }
	\label{fig:2}
\end{figure}

The wavelength dependence of the MOKE signals carries important information on the magnetization dependent optical transitions in CGT, allowing us to gain insight on the spin/orbital-dependent electronic structure and many-body phenomena, e.g. excitonic states.
We determine the MOKE spectrum by measuring magnetization curves at photon energies ranging from 1.13 to 2.67 eV (wavelengths from 465 to 1100 nm), and extracting $\varepsilon_A$ and $\theta_A$ for each of them.
At the same time we measure the reflectivity of the sample.
Fig. \ref{fig:2} shows the resulting spectrum of the reflectivity (a), the derivative of reflectivity with respect to photon energy (b), and the Kerr rotation and ellipticity (c).
We observe several features in both the reflectivity and MOKE signals, indicating a variety of optical transitions.
Around 1.43 eV, the MOKE spectrum shows a prominent feature reminiscent of an optical resonance.
This is also visible in the reflectivity spectrum, showing a feature in the same energy range, which is clearer when looking at the derivative of the reflectivity.

\section{Calculated Kerr Spectrum}
\begin{figure}[htbp]
    \centering
    \includegraphics[width=0.5\columnwidth]{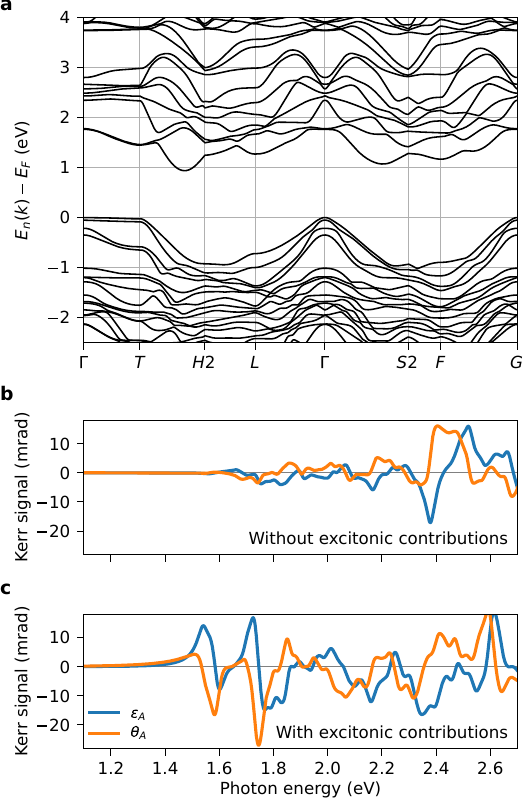}
    \caption{\textbf{Ab-initio bandstructure and Kerr signals.} {\textbf{a} Hybrid functional band structure of bulk CGT calculated along high-symmetry lines of the Brillouin zone. The Kerr angles (orange) and ellipticities (blue) calculated using the time-dependent Hartree-Fock approximation without (\textbf{b}) and including (\textbf{c}) the effects of excitons. See Methods for details.}}
    \label{fig:bs_bse_kerr}
\end{figure}

To gain further insight into the transitions and interactions contributing to the MOKE spectra, we compare our results to theoretical calculations.
Previous \textit{ab-initio} density-functional theory (DFT) studies \cite{Fang2018} taking only local Coulomb interactions within a mean-field GGA+U scheme into account and approximating the optical conductivity tensor within an independent single-particle picture, do not allow to reproduce our measured feature around 1.43 eV and only approximate the one around 2.6 eV.
We expect this to be a result of neglected long-range Coulomb interactions in these DFT studies, which have been shown to be important for determining the electronic band gap and magnetic interactions of CGT \cite{Menichetti2019} as well as for optical properties in other semiconducting layered magnetic materials \cite{Acharya2022}.
Therefore, we performed hybrid functional DFT calculations to account for the enhanced Coulomb interaction effects to the single-particle band structure of semiconducting CGT and subsequently evaluated the optical conductivity tensor within a time-dependent Hartree-Fock approach to include excitonic effects in the MOKE spectrum (see Methods for details).
The resulting band structure calculations are depicted in Fig.~\ref{fig:bs_bse_kerr}a.
As a result of the long-range Coulomb interactions, the indirect gap is enhanced to about 1 eV.
We find that, in agreement with previous results, our calculated magneto-optic Kerr spectra neglecting excitonic contributions to the optical conductivity do not reproduce the main features around 1.4 eV and 1.7 eV obtained experimentally.
However, when excitonic contributions are taken into account we observe features at 1.6 eV and 1.7 eV which are remarkably similar to the ones observed in our experiments.
As the hybrid functional and time-dependent Hartree-Fock calculations are only approximations, which cannot account for all (screening) details of the Coulomb interactions, we understand the quantitative difference between the measured and calculated magneto-optic Kerr spectra to originate from these approximations and expect even better agreement upon utilizing \textit{ab-initio} $GW$+BSE calculations.
The agreement between our experiments and calculations provides strong support for the major role played by excitonic resonances on the magneto-optic spectra of CGT.

\section{Electrostatic Control of the Magneto-Optical Efficiencies}
The excitonic features in the magneto-optical spectrum of CGT are highly tunable by electrostatic gating.
By exploiting the dual-gated structure of our devices we are able to independently control $\Delta n$ and $\Delta D$ and obtain the dependence of the MOKE signal measured during magnetic field sweeps, such as the ones shown in Fig. \ref{fig:1}c.
As discussed before, the saturation level of the MOKE signal is affected by $\Delta n$.
In literature, these changes on the absolute value of the magneto-optical signals are often associated to a change in the saturation magnetization.
Here we show that this is not the case and that the change in $\varepsilon_A$ and $\theta_A$ are, for a large part, caused by a change in the strength of the Kerr effect, i.e. the magneto-optical efficiencies.
A change of the saturation magnetization should result in the same relative change of both $\Delta \varepsilon_A$ and $\Delta \theta_A$.
Nonetheless, this is not what we observe.
Our measurements clearly show different behavior for $\Delta \varepsilon_A$ and $\Delta \theta_A$, with, for example, $\Delta \varepsilon_A$ decreasing and $\Delta \theta_A$ increasing with an increase in $\Delta n$.
Therefore, our measurements demonstrate the control of MOKE efficiency by the gate-induced charge carriers.

\begin{figure} [htbp]
	\centering
	\includegraphics[width=0.75 \columnwidth]{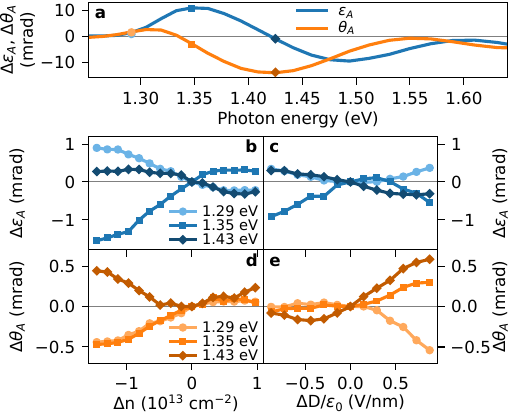}
	\caption{\textbf{MOKE dependence on $\Delta n$ and $\Delta D$.}
    \textbf{a}, MOKE spectrum around the excitonic resonance at 1.43 eV.
    Absolute changes of $\varepsilon_A$ (\textbf{b} and \textbf{c}) and $\theta_A$ (\textbf{d} and \textbf{e}) with $\Delta n$ and $\Delta D$ measured at photon energies of 1.29 eV (light colors), 1.35 eV (medium colors) and 1.43 eV (dark colors).
    }
	\label{fig:3}
\end{figure}

To obtain a clear picture on the electric control over the magneto-optical efficiencies, we focus on the $\Delta n$ and $\Delta D$ dependence of $\varepsilon_A$ and $\theta_A$ around the excitonic feature near 1.43 eV  (see Methods for more details).
A close-up of the resonance feature in the MOKE spectrum is depicted in Fig. \ref{fig:3}a, highlighting three representative photon energies of which the gate dependences are presented in the other panels.
Figs. \ref{fig:3}b and c show the absolute changes in $\varepsilon_A$ ($\Delta \varepsilon_A$) as a function of $\Delta n$ and $\Delta D$, respectively.
The simultaneously measured absolute changes in $\theta_A$ ($\Delta \theta_A$) are shown in Fig. \ref{fig:3}d and e.
Consistent with the data presented in Fig. \ref{fig:1}a, Figs. \ref{fig:3}b and d show that $\Delta \varepsilon_A$ and $\Delta \theta_A$ change by different ratios for different values of $\Delta n$.
A similar behavior is found for a change in $\Delta D$, as shown in Fig. \ref{fig:3}c and e, and for different photon energies, ranging from 1.26 eV to 1.61 eV, which are presented in Supplementary Fig. 2 and 3.
In general, for all these photon energies, the dependence of $\Delta \varepsilon_A$ on $\Delta n$ and $\Delta D$ is very different from the dependence of $\Delta \theta_A$.
Additionally, we observe different behaviors of the MOKE signals as a function of $\Delta n$ and $\Delta D$ for different incident photon energies.
These observations reinforce our conclusion that mainly the strength of the Kerr effect, and not the saturation magnetization, is affected by $\Delta n$ and $\Delta D$.

Our results show that $\varepsilon$ and $\theta$ can be modulated by up to just over 1 mrad and 0.5 mrad, respectively, by controlling $\Delta n$ and $\Delta D$.
These changes occur at 1.35 eV and correspond to 15$\%$ of the total signal. At 1.30 and 1.32 eV, the signal is smaller but changes by as much as 65$\%$   (see Supplementary Fig. 3).
A possible explanation for the dramatic changes in $\varepsilon_A$ and $\theta_A$ with $\Delta n$ can be given by a change in the strength of the excitonic resonance due to charging effects, and role of charged excitons for the magneto-optical spectrum.
This is reminiscent to the control of exciton resonances in 2D semiconductors, which shows a strong dependence of the oscillator strength and peak position of (charged) excitons with charge carrier density \cite{RevModPhys.90.021001}.
We note that an explanation of the dependence of the magneto-optical signals with $\Delta D$ requires a more in-depth understanding of our device.
The finite thickness of our CGT flake can lead to an asymmetric doping at the opposite surfaces for a given non-zero $\Delta D$.
This effect, in combination with the non-monotonic dependence of the magneto-optical signals with $\Delta n$ can lead to a more complicated behavior of the signals with changes in $\Delta D$.
For these reasons, we refrain from making any assumptions on the possible mechanisms for the modulation of the signals with $\Delta D$.

In addition to the saturation level, we have also determined the dependence of the coercive field $H_C$ and the nucleation field with changes in $\Delta n$ and $\Delta D$.
Since these are independent of the magnitude of the MOKE signal, the values of $H_C$ and $H_n$ can be accurately obtained by MOKE measurements.
We observe that both $H_C$ and $H_n$ depend on $\Delta n$, and are significantly more sensitive to hole doping ($\Delta n > 0$) than electron doping ($\Delta n < 0$).
The nucleation field is changed by $\sim$30\%, from $\mu_0 H_n = 4.0$ to 2.7 mT, when $\Delta n$ is varied from -1.5 to $1.0 \times 10^{13}$ cm$^{-2}$.
The coercive field is doubled from 0.3 mT to 0.6 mT by changing $\Delta n$ from -0.4 to $1.0\times10^{13}$ cm$^{-2}$, while it stays mostly unchanged when $\Delta n$ is decreased to values below $-0.4\times10^{13}$ cm$^{-2}$.
The change in $H_n$ and $H_C$ is likely to be caused by a change in the magnetocrystalline anisotropy, as pointed out by earlier works \cite{Verzhbitskiy2020, Hendriks2024}.
The complete discussion of these results are presented in the Supplementary Information.

\section{Conclusions and Outlook}
The importance of excitonic contributions to the MOKE spectrum of CGT we show here is crucial for the further understanding of magneto-optical measurements in this material.
Moreover, our demonstration of electrical control over the magneto-optical strength around the excitonic resonance provides a viable route for the manipulation of light polarization in magneto-photonic devices, where information stored in a magnetic material can be deterministically converted into changes in light polarization.
We envision that a reduction of the thickness of the CGT flake, exploiting thin-film interference effects \cite{Hendriks2021}, or using other van der Waals magnets with stronger excitonic features \cite{Molina-Sanchez2020}, should lead to a more dramatic gate modulation of light polarization.
The presence and electrostatic control over the excitonic resonances shown here also provide a key piece of information for the understanding of opto-magnetic effects.

\section{Acknowledgments}
We thank J. G. Holstein, H. Adema, H. de Vries, A. Joshua and F. H. van der Velde for their technical support.
This work was supported by the Dutch Research Council (NWO) through grants STU.019.014 and OCENW.XL21.XL21.058, the Zernike Institute for Advanced Materials, the research program “Materials for the Quantum Age” (QuMat, registration number 024.005.006), which is part of the Gravitation program financed by the Dutch Ministry of Education, Culture and Science of Education, Culture and Science (OCW), and the European Union (ERC, 2D-OPTOSPIN, 101076932). Views and opinions expressed are however those of the author(s) only and do not necessarily reflect those of the European Union or the European Research Council. Neither the European Union nor the granting authority can be held responsible for them.
The device fabrication and nanocharacterization were performed using Zernike NanoLabNL facilities.

\section{Methods}
\subsection{Sample Fabrication}
The device used here is the same as the one shown in Ref. \cite{Hendriks2024}.
The device fabrication goes as follows.
The thin flakes of 2D materials are obtained by mechanical exfoliation from commercially available bulk crystals (HQ graphene) on an oxidized silicon wafer with 285 nm oxide thickness.
For CGT this is done in an inert nitrogen atmosphere inside a glove box with water and oxygen content lower than 0.5 ppm.
The flakes are selected based on optical contrast, and stacked using a dry transfer van der Waals assembly technique \cite{Zomer2014}.
With a stamp consisting of a polycarbonate film supported by a piece of polydimethylsiloxane, we pick up the flakes consecutively to build the stack on the stamp.
We start by picking up the top hBN (21 nm thick), followed by the CGT (10 nm thick).
Next, we pick up a thin graphite flake to make electrical contact with the CGT. 
It only contacts a corner of the CGT to not be covering the part that is probed by the laser during the MOKE measurements, and it extends beyond the top hBN flake.
Afterward, we pick up the bottom hBN (20 nm thick) followed by a thin graphite flake that will act as the back gate.
The stack is then transferred to a transparent substrate (fused quartz).
To finish the stack, we transfer a thin graphite flake on top of it to act as a top gate.
Finally, the three graphite flakes are contacted by Ti/Au (5/50 nm) electrodes, which are fabricated using conventional electron-beam lithography and thin metallic film deposition techniques.

\subsection{Magneto-Optical Measurements}
The light source of our MOKE setup is a supercontinuum laser (NKT Photonics SuperK Extreme), from which we select a particular wavelength using an acousto-optic tunable filter (NKT Photonics SuperK Select).
The light passes through a beam expander and is directed towards a Glan Taylor polarization filter, transmitting horizontally polarized light, followed by a photo-elastic modulator (PEM).
The PEM has its optical axes set at 45$^\circ$ with respect to the polarization filter, and modulates the polarization of the light at 50 kHz with a retardance amplitude of 2.62 rad (0.417 $\lambda$).
Afterwards, the light is reflected by a 50/50 non-polarizing beam splitter towards the sample.
The light hits the sample at normal incidence, and is reflected back towards the beam splitter.
The transmitted part enters the detection stage, consisting of a quarter wave plate followed by a Glen-Taylor polarization filter and an amplified photodiode.
The quarter wave plate and the polarization filter are adjusted to compensate for the change in polarization caused by the optical components of the setup.
The Kerr ellipticity and rotation are then proportional to, respectively, the first and second harmonic of the output voltage of the photodetector, which are measured simultaneously using two lock-in amplifiers.
The average output signal of the photodiode, which is proportional to the reflectivity of the sample after subtracting the background signal, is also recorded.
From this data, the Kerr rotation and ellipticity are calculated \cite{Schellekens2014}. 

The sample is mounted in a flow cryostat (Janis ST-500), and is kept at a temperature of 10 K and pressures around $5\times 10^{-7}$ mBar for all measurements.
An electromagnet placed around the sample is used to generate an out-of-plane magnetic field.

\subsection{Gate Dependence Measurements}
We determine the gate-induced change in saturation level of the MOKE signal as follows.
First we ramp the magnetic field to +31 mT to saturate the magnetization of the CGT, and perform three gate sweeps while measuring the MOKE signal.
Then we ramp the magnetic field to -31 mT to saturate the magnetization in the other direction, and again perform three gate sweeps.
The MOKE saturation level is calculated by taking the average difference between the gate sweeps at +31 mT and -31 mT.
To obtain the gate-induced change in the MOKE saturation level, the value at zero gating is subtracted. 
For converting the applied gate voltages to values of $\Delta n$ and $\Delta D$, we use a parallel plate capacitor model with 20 nm thick hBN as dielectric for the bottom gate and 21 nm thick hBN for the top gate, taking $\varepsilon_{hBN} = 3.0 \varepsilon_0$ \cite{Pierret2022, Yang2021hBN, Hendriks2024}.

\section{Ab-initio Calculations}
All \textit{ab-initio} calculations are performed within the Vienna Ab-Initio Simulation Package~\cite{Kresse1993, Kresse1996} (VASP) applying 6$\times$6$\times$6 $k$-grids and an energy cut off of approx. 227 eV. The lattice constants and atomic positions are taken from the Materials Project for CrGeTe$_3$ (mp-541449) from database version v2023.11.1.

The electronic band structure is evaluated using a hybrid functional with the range-separation parameter set to 0.1 $\AA^{-1}$ and a fraction of $0.35$ of the exact exchange. These parameters were chosen to mimic more reliable QSGW band structure results by Ke~\cite{Lee2020} with a special focus to reproduce the QSGW band gap of about 1eV. In all calculations we take the spin-orbit coupling within the ferromagnetic phase into account. The band structure is extracted via Wannier interpolations using Wannier90 \cite{Mostofi2008}.

The retarded dielectric tensor $\varepsilon_{\alpha\beta}(\omega)$ is calculated by solving the Casida/Bethe-Salpeter equation with the same exact-correlation kernel $f_{xc}$ as utilized in hybrid functional calculations for the single-particle properties. This is an approximation to conventional $GW$+BSE calculations, which is here equivalent to time-dependent Hartree-Fock calculations. These calculations take both RPA-like bubble and ladder diagrams into account. The latter introduces the effects of electron-hole pairs (excitons) to the dielectric tensor. For these calculations we use the upmost 16 valence and the lowest 12 conduction bands and apply a frequency cutoff of 7 eV. For comparisons we also calculate the dielectric tensor within the random phase approximation, which excludes the effects of excitons.

\subsection{Calculation of MOKE spectrum}
The Kerr angles and ellipticities are calculated in the polar MOKE geometry from the frequency-dependent complex dielectric tensor $\varepsilon_{\alpha\beta}(\omega)$ obtained on different levels of approximations. Specifically, we use the standard RPA method based on single-particle DFT calculations, and a more sophisticated time-dependent hybrid functional DFT framework, which is supposed to capture the excitonic effects. The expressions for the Kerr angles are ellipticities read \cite{MOKE_book}
\begin{equation}
\epsilon_K = \frac{\mathrm{Arg}(\chi)}{2} \text{   and   } \theta_K = \frac{1 - |\chi|}{1 + |\chi|},
\end{equation}
where $\chi = r_+/r_-$, and
\begin{equation}
r_{\pm} = \frac{  1 - n_s - (n_s\tilde{n}_{\pm}^{-1} - \tilde{n}_{\pm})  \mathrm{tanh}(i\omega h\tilde{n}_{\pm})  }{  1 + n_s - (n_s\tilde{n}_{\pm}^{-1} + \tilde{n}_{\pm}) \mathrm{tanh}(i\omega h\tilde{n}_{\pm})  }.
\end{equation}
Here, $h\approx 10$~nm~is the material thickness, $n_s = \sqrt{\varepsilon_s}\approx 2$ is the substrate refractive index, and $\tilde{n}_{\pm} = n \sqrt{1 \pm Q}$, where $n = \sqrt{\varepsilon_{xx}}$ is the refractive index of the magnetic material and $Q = i\varepsilon_{xy}/\varepsilon_{xx}$ is the Voigt parameter. In the limit $\omega h/c \ll 1$, the expression above simplifies to an expression frequently used in the context of 2D materials
\begin{equation}
\epsilon_K + i\theta_K = \frac{2n^2Q\omega h}{1- n_s^2},
\end{equation}
which is not fully justified for thick samples.

\section{Author information}
M.H.D.G. conceived and supervised the project.
F.H. built and tested the measurement setup, designed and fabricated the samples, performed the measurements and analyzed the data under M.H.D.G. supervision. 
A.N.R. and M.R. performed the theoretical calculations.
All authors discussed the data and contributed to the interpretation of the results.
F.H. and M.H.D.G. co-wrote the first version of the manuscript and all authors contributed to the final version of the text.

\bibliography{cgt_moke_spectrum}

\end{document}